# Cross-Domain Comparative Analysis of Digital Twins and Universalised Solutions

Guanyu Xiong, Haijiang Li*, Yan Gao

BIM for Smart Engineering Centre, School of Engineering, Cardiff University, Cardiff, CF24 3AA, United Kingdom

**Abstract**

Digitalisation is one of the main drivers of most economic sectors nowadays and the digital twin, as a reification of digitalisation for complex systems has attracted much attention from both academics and industry. There have been studies focusing on digital twins in a specific sector while there are few exercising insightful comparisons of digital twins from different domains. Considering the digital twinning is a cross-domain transformation, it is beneficial to establish the principles of universality and variation that can explain similarities and differences in any digital twins. This paper first delivers a comparative analysis of digital twins in five domains through a six-dimensional characterisation framework. Then, by departing from the correlations among the domain-specific DT development, a cross-domain Digital Twin Platform-as-a-Service (DT-PaaS) is proposed to universalise the common process, tools and applications, meanwhile being inclusive of variations of every digital twin instance. As a centralised data, modeling and service platform, it is expected to break the barriers between domains by enabling the cross-domain digital twin data sharing, interoperability and development synergy and tackle some complex global challenges such as climate challenge, net zero, pandemics, etc.

## 1  Introduction

Digitalisation, often regarded as the fourth major innovation cycle in human history, is changing almost all aspects of economic activities [1]. Of almost all the major technology trends – from artificial intelligence to autonomous vehicles, data-centric and digitalisation are becoming the main drivers to shape the future world. One reification of digitalisation that is increasingly being applied to engineering systems is digital twinning which can empower intelligence to non-living assets, thus enabling engineering systems to understand, predict and optimise their own performance.

The development of digital twin (DT) has been a cross-domain effort. The DT concept was originated from the aerospace domain in the 1970s during the Apollo 13 program to enable engineers on the ground control the vehicles in the aerospace [2] and then adopted as a major vision of the industry 4.0 development in the manufacturing domain for processes development and life cycle management [3]. Following other industries and fields such as integration of DTs for real-time monitoring of urban mobility and sustainable development within smart cities, and medical resource management and precision medicine within the healthcare.

However, the existing research on DT tend to focus on a specific domain. Though some of them conducted comparative study, they tend to use an advanced DT domain, mostly manufacturing, as a reference for comparison against their interested domains. For example, Mylonas et al. [4] surveyed DTs applications in smart cities against those in Industry 4.0. Taylor et al.[5] presented review of the development of digital twins in the manufacturing and maritime domains. Davila Delgado & Oyedele[6] summarised DT features of manufacturing and discussed their applicability for built environment. There are also studies examined DT of multiple domains but the comparison is rather brief or unstructured. Madubuike et al. [7] examined DT applications in various industries with a focus on construction, the aspects of comparison are the usage and benefits only. These studies made positive contribution to their own domains, but simple one-to-one comparison may struggle to uncover the wider cross-domain relevance.

On another aspect, most DT review studies are based on other DT literatures while neglect tracing the sources of the terminology or discussing the principles. As digital twinning is an interdisciplinary study, the advancement of DT has utilised concepts from a range of engineering subjects, such as complex system engineering, software engineering, modeling engineering, etc. So it is considered necessary to dive deep enough to gather groundings that DTs are developed upon and discover insights that can provide long-term vision for the future of DTs.

Generally there are two fundamental questions (FQ) that are dominating the DT related research [8][9][10][11]: FQ1 - What is the concept of DTs and; FQ2 - How to implement DTs. FQ1 has been overwhelmingly discussed in publications and it has been fundamentally agreed that the DT is a high-fidelity virtual replica of a physical object, enabled by bidirectional and real-time data communication, to achieve intelligent decision-making. Therefore, this paper focuses on FQ2 by providing a characterisation framework which is summarised from the key perspectives from the implementation process of a DT, and this framework shall be applicable to any DT, regardless of the domain.

A cross-domain comparison required two steps: establish a universal ruler of which metrics are based on the procedures involved in the creation of a DT and then use this ruler to measure DT from related domains. There are papers attempted summarising characteristics of DTs [8] and producing DT descriptive models [4]. These models are proposed based on review study of several DT domains and established commonalities. While the underlying reasons for the presence of the commonalities and differentiation are not investigated. The Gemini Principles [12], a recognised DT framework, can satisfy most requirement described above. It introduces an outcome-oriented approach to DTs by defining generic objectives, while how to achieve these objectives - the implementation of DTs, has not been specifically discussed. Hereby it is argued that characterisation framework describing the key procedures for developing DTs and a cross-domain comparison based on this framework are required to bridge the identified research gaps.

The rest of this paper is organised as follows: Section 2 presents the methodology of this study, including the rationales for the selection of domains and the synthesis of the six-dimensional DT comparative framework. Archetypes of DT instances are selected for domain and features identified in the comparative framework are extracted for are referenced in the domain-specific analysis of Section 3. Observations and underlying reasons that universalise and differentiate DTs are discussed in Section 4. Finally, a summary and conclusion are drawn in Section 5.

## 2 Methodology

With the research gaps and research significance discussed in Section 1, the overall research objective - how to conduct a cross-domain DT comparison, was destructed into step-by-step procedure described by the following research questions:

- RQ1: what are the most discussed DT domains and what domains shall be selected for the comparative analysis?
- RQ2: How to establish a framework eligible to compare DT of any domain?

- RQ3: What are the characteristics of DTs in each domain, with reference to the comparative framework?
- RQ4: What are the similarities and differences in the characteristics of the DT of each selected domain?
- RQ5: How to utilise the similarities and pave the way for the synergy of the cross-domain DT development?

The five research questions and the corresponding research methods were formed into the methodology illustrated in Fig.1.

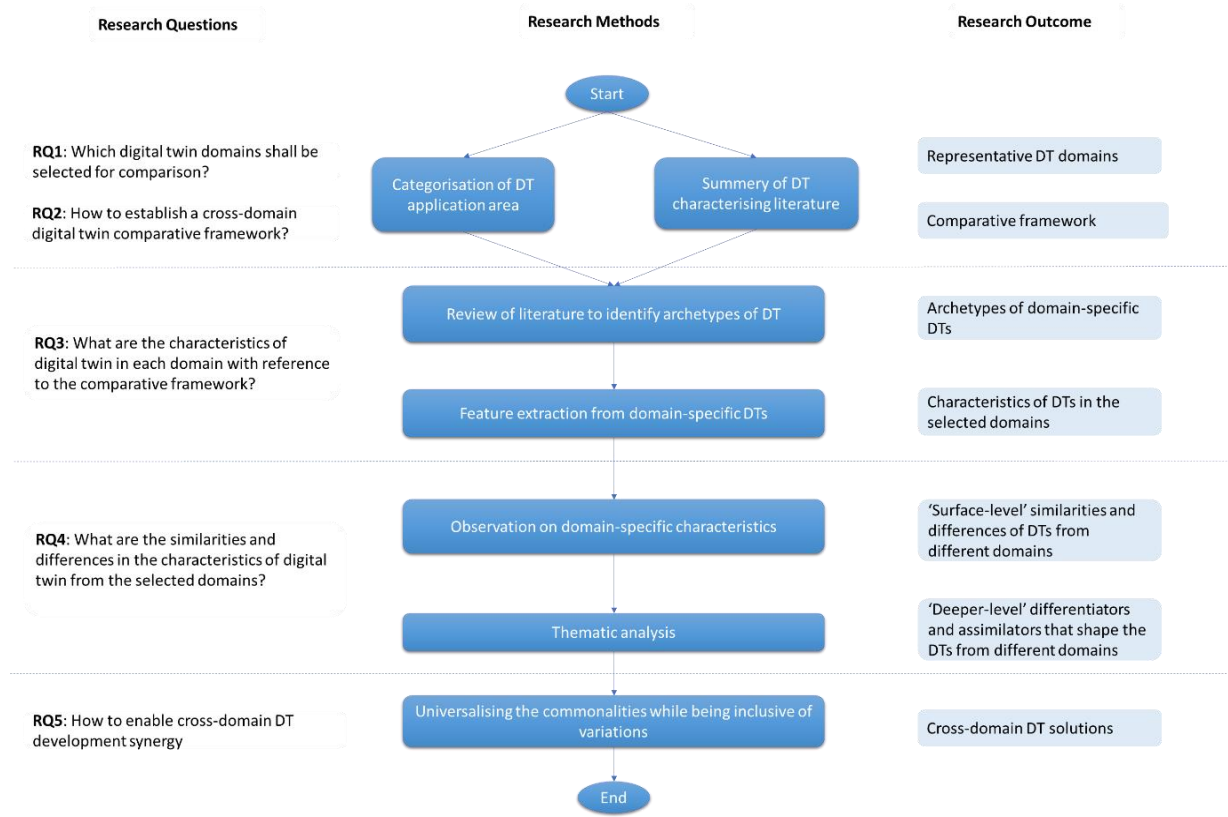

Figure 1: Methodology Diagram

To answer 'RQ1: How to select domains for the comparative analysis of DTs?', authors first identified that the domains selected shall be representative and inclusive of major DT applications. To achieve this goal, all the DT-related publications from Scopus and Web of Science were classified with the filtering tool to provide two lists of DT subjects- one for each database. Then a classic Three Sector Model was leveraged to aggregate and refine the filtered subjects into three groups. One or two representative domains of each group were selected and introduced to the in-depth analysis and cross-domain comparison.

To establish a DT comparative framework for RQ2, the key attributes of DTs that can be used for cross-domain comparison were conceptualised to three levels of categories. So that every DT use-case can be described with higher clarity and accuracy.

In the domain specific DT analysis required by RQ3, some typical DT use-cases were selected from each interested domain and the comparative framework was applied to assess these use-cases. The characteristics of these use-cases were extracted and correlated to underlying reasons of the phenomena, normally implied by the demand or nature of the domain.

As the basis for RQ4, the differences and similarities of DTs from different domains were summarised and analysed using a thematic analysis method - Principles of Variation and Universality [13]. This method of comparative analysis starts from 'surface-level' similarities or differences and implies that they potentially could be explained by 'deeper-level' assimilators or differentiators, thus forming an explanatory theory.

With the throughout and insightful understanding of DTs of various domains underpinned by the six-dimensional framework supplemented with the explanatory theory, corresponding universal DT solutions were proposed and integrated as a unified cross-domain Digital Twin Platform-as-a-Service (DT-PaaS), as an answer for RQ5.

### 2.1 Selection of Domains

To estimate the presence of DTs in every economic sector, a cluster of publications was obtained with the search query "digital twin" from the academic publication databases Scopus and Web of Science. And then the filtering tools of the two databases were applied to classify the collected cluster based on subject area.

In economics, production chain is used as an analytical tool to understand the nature of the production process. According to the Three Sector Model [14], production chain can be divided into three classes: primary industry where raw material are extracted, secondary industry which involves the transformation of raw materials into goods and tertiary industry which delivers services to customers. For each industry, representative sectors are selected for review of DT research and application following the procedures illustrated in Figure 2.

Two clusters of research domains obtained from Scopus and Web of Science were refined and condensed, and then combined according to the Three Sector Model. Then the domains were ranked based on number of publications and selected based on the presence in the research publications.

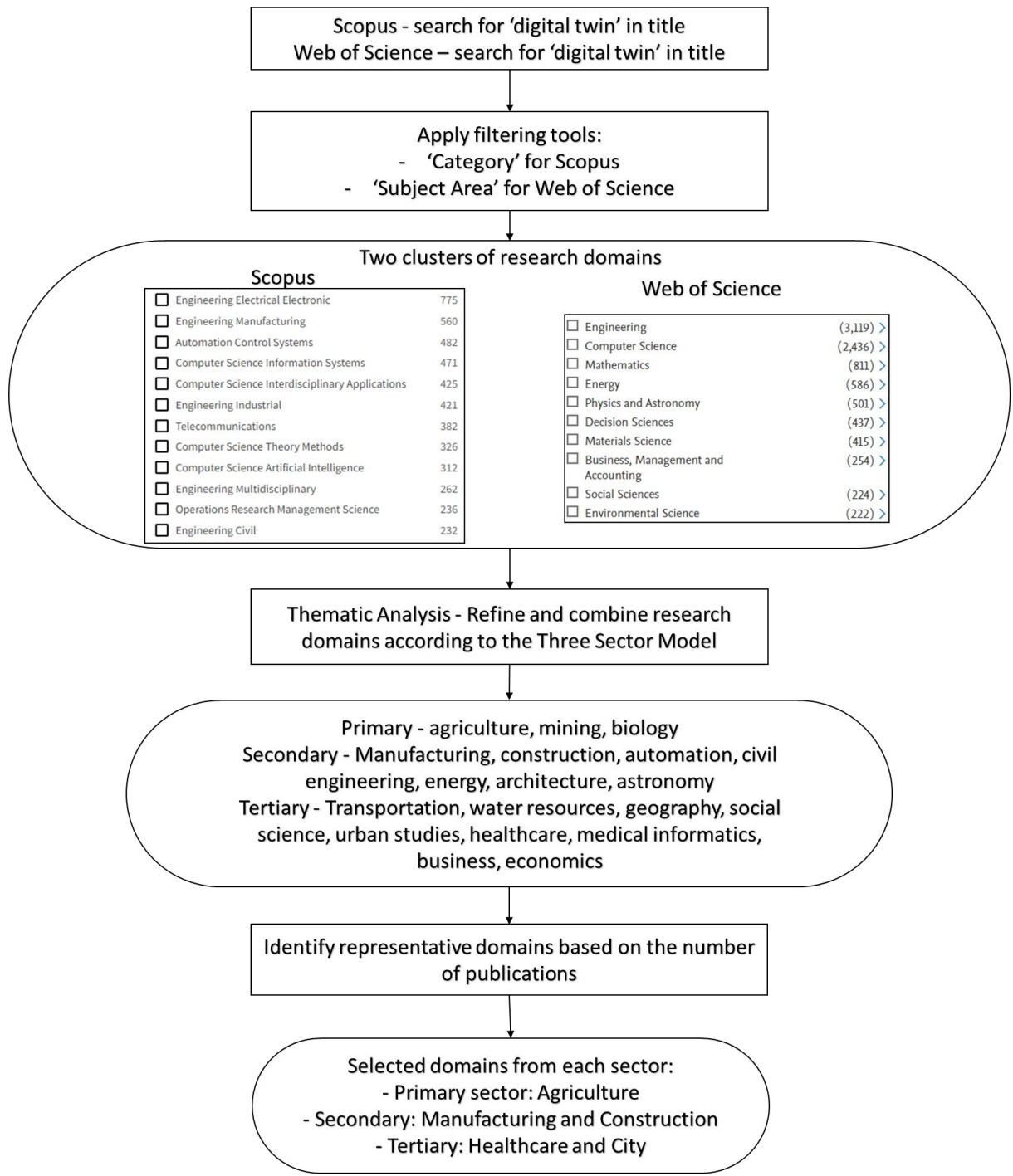

Figure 2: Selection of digital twin domains for comparative study

## 2.2 Synthesis of Comparative Framework

The exercise of DTs comparison on multiple domains can be considered as a comparative analysis. According to the methodology described in [13], as the purpose of the study is to establish that every concept, implementation and use-case of DTs follow the same rule, a universal frame of reference of the following criteria shall be determined:

- inclusive of all the commonly agreed core elements of DTs;
- the elements shall present underlying causal patterns;
- universally applicable to DTs of any domains, scales, use-cases, etc;
- include sufficient details while being reasonably simplified for the conciseness of the comparative study

Currently there is no clearly defined criteria to compare DTs of the selected domains, therefore a framework for comparison is essential where the measurement of DT need to be stated. In addition, as the appearance and behaviour of DTs are not directly measurable, a process of operationalisation is also needed to define the parameters and criteria for comparison. With this framework, features of typical DTs of the selected domains can be extracted and compared. Hence, the DT comparative framework is synthesised from the following steps: conceptualisation of DT attributes, characterisation of DT attributes and operationalisation of the attributes for measurement and comparison.

Relevant work has been reviewed against the above criteria. The sources take input from both empirical evidences observed by industry experts and systematic literature corpus derived by academic researchers. The result is summarised in Table 1.

| References | Domains | DT Characterising Frameworks | Comments on frameworks |
|---|---|---|---|
| [8] | Multiple domains | 1) Application context<br>2) Life-cycle phases<br>3) Functions<br>4) Architecture<br>5) Components/technologies | - Defined where, when, why and how to develop a DT<br>- Physical object and digital modeling are not addressed |
| [9] | Manufacturing, energy, aerospace, automotive, agriculture, healthcare | 1) Industrial sector<br>2) Purpose<br>3) Physical reference object<br>4) Completeness<br>5) Creation time<br>6) Connection | - Provide perspective on the application scenarios of DT across industries<br>- DT technical development details are not included |
| [10] | Multiple domains | 1) Goals<br>2) User focus<br>3) Life cycle focus<br>4) System focus<br>5) Data sources<br>6) Data integration level<br>7) Authenticity | - For application-oriented DT applications and universally valid in all DT related domains<br>- All the key elements are covered but lacks details and the links on the elements |
| [11] | Manufacturing | 1) Physical entity/virtual twin<br>2) Physical/virtual environment<br>3) State<br>4) Metrology<br>5) Realisation<br>6) Twining rate | - A complete conceptual description of the DT<br>- Some implementation tools and technology are described |

| | | 7) Physical-to-Virtual connection<br>8) Virtual-to-Physical connection<br>9) Physical/virtual processes | |
|---|---|---|---|
| [15] | Based on Manufacturing domain | 1) Purpose<br>2) Data input<br>3) Data link<br>4) Synchronisation<br>5) Interface | • Possible to be used as reference to measure the DT development progress<br>• Digital model and application/services are not introduced |
| [16] | Multiple domains | 1) Scope of physical entity<br>2) Feature of physical entity<br>3) Scope of virtual entity<br>4) Form of data communication<br>5) User-specific output/values | • For cross-industry classification and development of applications within the concept of the DT<br>• Major DT elements are covered, can be used as basis for creating a more detailed framework |
| [17] | Multiple domains | 1) Application areas<br>2) Federation<br>3) Layering<br>4) Spatial scale & resolution<br>5) Temporality & resolution<br>6) Lifecycle stage<br>7) DT actors & asset stakeholders | • Enable decision-makers to articulate the DT user requirements<br>• involve supply and delivery of a complex DT by various multiple parties<br>• Technical elements are not fully covered |
| [18] | For manufacturing systems | 1) Physical entity<br>2) Virtual model<br>3) Service<br>4) Data<br>5) Connection | • Emphasis on functions and practice, clear causal links are presented between DT elements<br>• Can be further extended for cross-domain comparison |

Table 1: Digital Twin Characterising Frameworks

By summarising the above DT characterising frameworks, it is firstly noted that characteristics of DT can be generally divided into two large groups: abstractions and concepts related and implementation-focused tools and techniques. These phenomena also match with the two fundamental questions FQ1 and FQ2 described in Section 1. Therefore, it is proposed that the frame of reference shall be primarily divided into two categories - conceptualisation and implementation, so that DTs of any domains can be measured in conceptual development and technical development.

Many publications [9][11][16][17] state that the starting point of digital twinning is the physical entity and the purposes shall be clear in the beginning, as highlighted by The Gemini Principles[19]. The purposes of digital twinning are inclusive of the goals to be achieved[10], the values to be created [16] and the application to be realised [17]. To bridge the starting point and purposes, bi-directional and synchronised data link[9][11][15][16] shall be introduced, alongside with some other elements (users [10] [16] [17], life-cycle stage [8] [10] [17], etc), to be packaged into the system architecture, as a presentation of the whole picture of the conceptualised DT.

After answering the RQ1: what is the concept of DT, next comes the RQ2: how to implement DT, which is the process of turning the concept of a DT into reality via various skills, tools and technologies. Three steps were identified for the technology-oriented implementation. To begin with, the behaviours of the physical entity are captured in the form of data via technologies such as sensors, cameras and scanners. To interpret the data, certain levels of computing power are required so the data is transmitted to another venue, normally the cloud. There might be reduction in the complexity of data during the communication, but the nature of the data remains unchanged until the next step where data is reconstructed to form models that can mimic the operation of the physical object. While setting up the modeling of the physical object might be a milestone but it is not the destination. As the final step, different levels of services are delivered to the users to meet their demands defined in the purposes.

Thus, a six-dimensional comparative framework of DTs is shown in Fig.3, it describes a DT from characteristics in the conceptual development and technical implementation. As justifications from Section 2.2.1 to Section 2.2.6, it is extended to multi-hierarchical to include representative categories and sub-categories based on the DT studies from different domains.

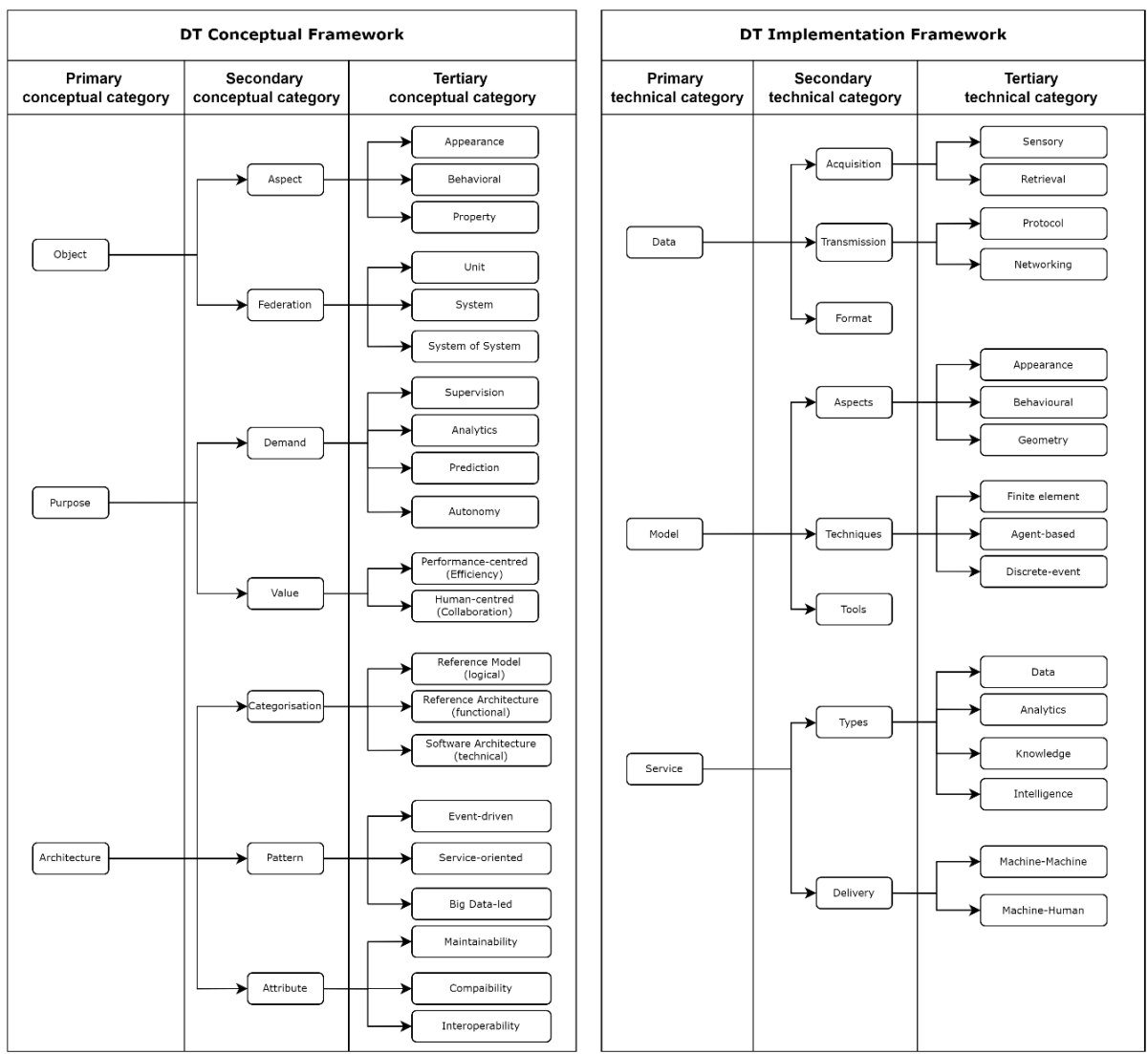

Figure 3: Six-Dimensional Cross-Domain Digital Twin Comparative Framework

### 2.2.1 Twinning Objects

A twinning object is the entity that the DT is created against, so the identification of the twinning object is the first question to consider when creating a DT. The initial twinning objects are entities physically exists and human-made, such as a product [20] or a factory[21], while in recent years the interest has grown into natural entity such as animal[22], climate system[23] and those abstractly exist, such as an enterprise DT which replicates an organization's operation that does not physically exist [24] [25]. The commonality of these entities is the existence in the real world, either physically or abstractly, so that they can be digitally twinned in the virtual world of the cyber space. Hence, though some papers refer to the twinning object as 'physical twin', to be inclusive of the recent development and the wider applications, this study employs the term 'twinning object' for general applicability.

On another dimension, twinning of object only specifies some aspects of an object, such as geometrics, material or behaviours, depending on the type of information that the DT users are interested in. Hence it worths considering which aspect of the object is twinned and how it affects the twinning process.

Both industry and academia have established some practices to characterise twinning objects. Siemens [26] splitted the twinning object based on the stage of the product lifecycle - there are product twins for efficient design of new products, production twins for production planning and performance twins to capture, analyse, and act on operational data of manufactured products. IBM [27] defined twinning objects in terms of product magnification, starting from component twins, then asset twins to study the interaction of two or more components and system twins for entire functional system, finally process twins to reveal how systems work together. The biggest difference between these twins is the area of application. In a DT feasibility study for the Thames Tideway Tunnel scheme, Whyte *et al.* [28] proposed that various subsystem could be articulated in three levels for the purpose of digital twinning: asset, project and system of systems. The three levels could be divided according to different modeling techniques. Similarly, Rosen, Boschert and Sohr [29] proposed that production system shall be digitally twinned in three scales: product, process or system. Tao *et al.* [30] divided the entities into three levels according to function and structure, which are unit level, system level, and system of systems (SoS). Al-Sehrawy, Kumar and Watson [17] classified the federation of urban DT into sub-system, system and system of systems as the foundation for the different levels of systemic analysis, to enable inter-organisational collaboration and dissolve the infrastructure sectoral silos.

It is concluded that all twinning objects can be categorised as individual, assembly system, system of systems. On top of the system level, there can be system of systems for massive scale entity such as a city where systems are interconnected. The selection of twinning object depends on the research purpose and research object, different twinning object can determine the modeling techniques.

### 2.2.2 Twinning Purposes

Purposes of digital twinning are the reasons why a particular DT is created, they highlight the essential link between the DT apparatus and the applications, functions or use-cases that the DT is expected to achieve. DTs are built for various purposes, even with the same physical twinning object. The purpose of digital twinning determines the aspect of the physical asset to be digitised and the level of detail that the DT should encompass when compared with its physical twin [31]. Also the purposes could affect other key elements of DT, including design of system architectures [32], fidelity and dynamics of modeling[33], and the services to be delivered(Aheleroff *et al.*, 2021). So it is essential to define the purposes for which the DT is created before any subsequential design, development and implementation procedures.

There has been cross-domain effort in the review of digital twinning purposes. In the agriculture domain, Pylianidis *et al.* [35] categorised functions of DT into fundamental level with monitoring, user-interface and analytic components, enhanced level including actuator components to control, further enhanced level with simulation components and the final level with learning capability being able to find underlying mechanisms of systems. For the manufacturing domain, Wagg *et al.* [33] suggested a five-level capabilities hierarchy, supervisory, operation, to prediction, learning and finally autonomous management. Boje *et al.* [36] proposed a three-tier paradigm for construction DT platforms, namely monitoring platform, intelligent semantic platform and agent driven socio-technical platforms, according to the levels of intelligence and lifecycle integration. In the cities domain, Al-Sehrawy *et al.* [17] developed a multi-dimensional uses classification framework for city DTs, which primarily divides DTs uses into information mirror, communicate, analyse and control. On a cross-industry vision, Enders & Hoßbach [9] summarised the purposes of DT applications into three levels: monitoring, simulation and control on the basis of a systematic literature review into main DT application domains.

It is noted that the purposes of digital twinning within different domains share many commonalities, including real-time monitoring, fault analytics, simulation and prediction, optimisation, etc., and the classification of these purposes is mainly based on the level of capability and intelligence, which can be assessed from two aspects: 1) awareness and understanding of the situation and 2) timeliness of response to the situation. So, an analogy can be drawn as Figure 4 to match DT intelligence to human intelligence for better intelligibility.

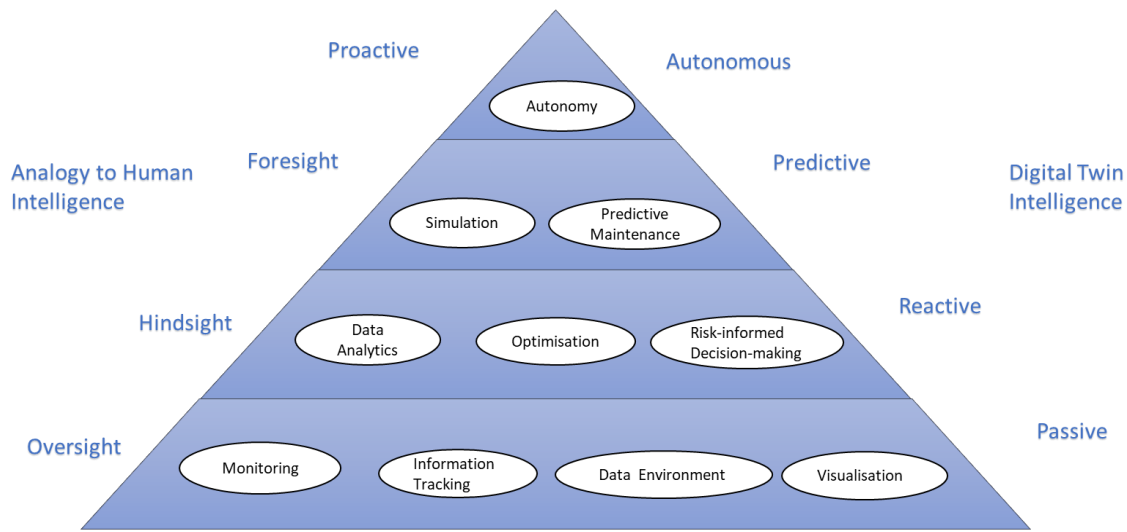

Figure 4: Digital Twin Intelligence as Analogy to Human Intelligence

The first two levels 'oversight' and 'hindsight' refer to the supervision of system status and decision-making support by data. Though DTs at these levels are aware of the situation and may be capable to analyse the root causes, but generally their performance are passive or reactive with delay in time. 'Foresight' level means simulation of the physical twin from models and data, so that prediction of future scenario can be exercised for pre-planning and targeted optimisation, but it still relies on human to close the control loop. A proactive and autonomous DT can achieve certain degree of autonomy by influencing its physical twin through actuators, so that lifecycle management of the asset can be incorporated with minimum human intervention. This DT intelligence hierarchy is applied as a universal scale for measuring the purposes of digital twinning.

On top of the basic demands, the values that DTs will create for organisation and society also motivate the adoption of this technology. The Gemini Principles [19] for DTs of built environment outlines the generic principles such as public good, openness, trustworthy, etc. Existing applications of DTs have been motivated by two main values: performance-centre and human-centre.

The DT intelligence described in Figure 3 such as optimisation, predictive maintenance and autonomy all emphasise performance improvement for better efficiency. While human-centred DT design has been increasingly discussed. Schrotter and Hürzeler [37] introduced DT of the City of Zurich for collaboration on urban planning and public awareness on climate change. Longo, Nicoletti and Padovano[21] propose human-centric manufacturing paradigm where manufacturing employees empowered with the knowledge about the manufacturing system are integral part of the factory and pave the way for production and business performance improvement opportunities. Full automation enabled by DTs could also introduce social impact. Elon Musk [38] argues that the control loop of manufacturing line shall be closed by staff to avoid "excessive automation". The unemployment created by automation could weaken the buying power and the business performance. However, the original intention of manufacturers' investment in automation and DT is for profitability.

### 2.2.3 System Architectures

The identification of twinning objects and twinning purposes can be a straightforward process but the activity to map them can be challenging. However, there exist established concepts and practices from other fields that can shed light on this challenge. In system and software engineering, after identifying the starting point (i.e. twinning object) and the objective (i.e. twinning purpose), the next step of the conceptualisation is to compose a system architecture that can enable the expected functionalities of the system[39]. To further conceptualise the architecture of DT, academia community have developed reference models [40] [41], reference architectures [34] [42] and some typical architectural styles such as multi-layered[43] and serviced-oriented [44] for digital twinning systems. The taxonomy is described by Len Bass and Paul Clements (2003) in the book Software Architecture in Practice where the completeness of software system architecture is defined in three levels: reference model, reference architecture and software architecture. Architectural styles are linked to system's quality attributes so they are the foundation for attribute-driven architecture design.

On the basis of these terminologies, there has been effort in academia to extract features of DT architecture and develop them into a classification framework. Eindhoven and Version [46] described a method to classify architecture framework for automotive software system based on the level of abstraction, starting from logical architecture to functional architecture and finally implementation architecture. Tekinerdogan and Verdouw [47] proposed a pattern-oriented approach for architecting DT-based agricultural systems with a catalogue of nine different architecture design patterns which address different use-cases (e.g. model, matching, proxy, monitor, control, autonomy, etc.). These patterns are essentially multi-layered architecture where each layer performs specific tasks, layers are added on for higher level of intelligence and complexity. Ghita, Siham and Hicham [48] presented RAMI4.0 (Reference Architecture Model for industry 4.0) inspired DT reference architecture and its variants for different industrial ecosystems including industrial internet of things, complex system engineering, cloud services, etc. These architectures are consisted of two identical primary layers and different functional elements in each layer. These elements can deliver different performance in different contexts such as data management, multi-agent interoperability, security management and functional suitability. Ferko, Bucaioni and Behnam [32] conducted a systematic mapping of proposed DT in literatures to architectural solutions, architectural patterns and quality attributes. This study concluded that most of DT architecture are built using a combination of the layered and service-oriented patterns and address maintainability, performance efficiency, and compatibility quality attributes.

Multi-layered architectures are referenced in most DTs implementation. Layered architecture allows modules or components with similar functionalities to be organised into horizontal layers, then each layer performs a specific role within the application[49]. In manufacturing domain, one basis for creating a DT architecture is the cyber-physical system (CPS) – the core concept of Industry 4.0. CPS is based on the ISA-95 architecture and a further developed into the 5C architecture which consists of five layers: connection, conversion, cyber, cognition, and configuration levels. Component-based or module-based architecture has been widely applied in manufacturing domain for crafting Cyber-Physical system[50], while service-oriented architecture are borrowed from recent development in computer engineering for multiple-services based DT implementation. [51] proposed a software-intensive DT architecture to achieve contextual-awareness and automatic self-management. But it is worth noting that DT architecture differentiates from a pure software system as the former include physical twin and a series of hardware for data acquisition, processing and delivery.

As a conclusion, the system architectures of DTs shall feature level of completeness, architectural patterns and quality attributes in the comparative framework.

### 2.2.4 Data

While twinning object, purposes and system architecture can explain how DTs can be conceptualised, it needs another technical framework to describe how implementation of DTs can be achieved. It is widely recognised that the basis of DT is data [10], the core of DT is modeling and then applications that are rooted on data and modeling are encapsulated as services to users. Hence the comparative framework classifies tools and technologies involved in the implementation of DTs into three dimensions: data acquisition and communication, creation of model and delivery of DT-enabled services.

The first step of DT implementation is to acquire the data capably of describing the conditions of the twinning object. The most proposed data acquisition method is automated collection in real-time via sensory technology including sensing of physical phenomenon and visual capture such as camera. These sensors can be mainly classified according to mechanisms, such as mechanical sensors, electronic sensors, chemical sensors and biosensors, and are integrated into the system for reliable data collection. Retrieval of historical and/or external data from database is also required in some applications. Newbery [52][52] describes a chemical process DT receives its live data from sensors attached to the physical asset, historical data from external databases, and humidity or temperatures from the national weather services.

Data acquisition is usually carried out through an elaborated sensor network at the edge under different scenarios, e.g., a city, a construction site, farmland, a factory, or even a person. The versatile sensors, e.g., gauges, high-solution cameras, scanners, QR tags and readers, are selected and applied according to DT services. Moreover, the sampling frequency (e.g., the data rate in A/D conversion for digital electronic equipment), which is highly related to the timeliness of the digital entity reflecting the physical entity, is determined according to DT services in terms of domain characteristics. For example, agriculture DT applications do not require high sampling frequencies, because crops have a long and slow growth process. In contrast, manufacturing applications, such as DT in 3D printing, usually ask for high sampling frequencies to guarantee the timely reflection of the fast manufacturing process in the DT, thereby ensuring the product quality. Another aspect is the placement of sensors (including amount, locations, etc.), which is strongly correlated with the fineness of the DT representing the physical entity. The optimal placement of sensors is always a hotspot in research since IoT emerged, and it varies across different domains according to specific domain knowledge, e.g., the optimal placement of accelerometers for structural health monitoring (SHM) need to consider structure elements, material, geometry, and selected algorithms, etc., while the placement of wearable sensors for healthcare depends on the medical knowledge and patient situation. Another research hotspot is energy-related technologies for sensors because a stable power supply is always a problem under many scenarios in implementation. There are three major directions for this research question: 1) renewable energy for sensor nodes, e.g., wind, solar; 2) sensors with ultra-low power consumption, which usually can keep running for several years based on batteries; 3) self-powered or energy-harvest sensors, e.g., self-power electrochemical biosensors and energy-harvest vibration sensors.

The transmission of data can be based on wired or wireless communication. Although wired networks have much greater data volume capability, wireless sensor nodes enables monitoring to areas or entities which used to be inaccessible. For example, wearable devices are usually moveable, requiring wireless communication, such as Bluetooth. Wireless communication can be classified according to transmission range, including short-range, mediate-range and long-range, or according to topological architecture, such as cellular or non-cellular. Another concern about the selection of appropriate communication technology is data rates and delays. Low data rates probably result in communication congestion when there are massive amounts of heterogenous data from ubiquitous sensors, while long delays will be a disadvantage for DT synchronisation. It is also worth noting that communication in a DT system is not based on a single technology but usually hybrid, e.g., in a smart city domain, NFC is responsible for resident identification; open Wi-Fi takes charge of public communication; NB-IoT or LTE-M takes care of infrastructure monitoring; optical fibre is used for the connection between gateways and cloud servers. For the industry with high-degree digitisation, there are already several standard protocols, e.g., in the smart manufacturing, OPC UA protocol is used for industrial telecommunication (machine-to-machine); Ethernet/IP used for industrial network; TCP/IP is suitable for network interconnection; UDP is a lightweight protocol, used to create low-latency and loss-tolerating connection; MQTT is also lightweight and used to publish and subscribe message.

Different from the widely applicable data transmission protocol, data formats and data models are specific to domains. In manufacturing industry, Exchange of Product Model Data (STEP) has been applied to store life-cycle product information in the context of DT [53] and AutomationML is proposed to store production monitoring information for manufacturing and product DT service [54]. As high-level models, STEP and AutomationML enables data storage and exchange between systems but relies on external middleware for data information extraction.

In the city domain, Geography Markup Language (GML) are applied in the modeling, transport and storage of geographic information for defining infrastructure and civil engineering activities [55]. But GML lacks asset management data capability to evolve into an operational system for DT applications.

Similarly, the Industry Foundation Class (IFC) widely adopted in construction industry is in need of an effectively leap from a static data model to a dynamic information sharing paradigm[36]. Wang *et al.* [56][57] and Chevallier, Finance and Boulakia [58] have developed extended IFC models to include static semantic structural description and dynamic time-series sensors data. Nevertheless, data enriched IFC does not support interoperability between interconnected DTs. A light-weight data solution that can enable cross-domain data integration, software integration and real-time synchronization is still needed.

As some more recent cross-domain effort, FIWARE [59][59] defines a series of Smart Data Models that not only define the technical data types and structure, but also provide a interoperable middleware between IoTs and software system. CDBB [60] proposed Foundation Data Models as the top-level ontology for any DTs and Reference Data Library describing common set of classes and the properties of domain-specific DTs.

### 2.2.5 Modeling

DT modeling is a process to replicate the interested aspect of the twinning object, such as physical geometries, properties, behaviours, etc. Modeling shall be constructed at the appropriate level of abstraction of the twinning object using appropriate modeling techniques [61].

To compare and contrast DT modeling, relevant literatures were reviewed and analysed to identify the dimensions of DT modeling that are potentially correlated to different DT domain. From the basic definition and concept of DT, Rasheed, San and Kvamsdal [62] suggested physics modeling is the workhorse at the design and construction phase of DT and it can further divided into three sub-types depending on the understanding level of physical phenomenon, namely resolved physics, modelled physics and observed physics. They can be combined with data-driven modeling to reduce the order of physics modeling thus providing a more flexible and expressive framework for rapid model adaptation with balanced interpretability [63]. Some wider frameworks were provided by Qi *et al.*[64] and Liu *et al.*[65][65], both studies stated that a DT model is assembled or fused by four types of sub-models (geometry, physics, behaviour and rule) of different characteristics and functions and model integration is the key to address the contradiction between simplified virtual model and complex behaviour of the twinning object. As an implementation of model fusion, Liu *et al.* [66] developed a DT mimic modeling method by merging geometric model, behaviour model and context model to deliver holistic information monitoring of machining process.

In addition to the types of models that a DT shall encompass, some modeling techniques have been specifically discussed due to their suitability in DT modeling. For example, Finite Element Analysis (FEA) is a typical physics-based modeling for assessing the behaviour of an assembly under physical effects. For a complex system in which many parts interact and influence each other, agent-based modeling is normally applied. Discrete event modeling is for systems that can be decomposed into a set of logically separate processes that autonomously progress through time[67]. The construction of the modeling is carried out in modeling tools, of which mostly are desktop-based. While some web-based modeling tools are developing rapidly in recent years which feature light-weight, open-source for better interoperability with other software systems. Some examples are Xeokit for building information modeling and Cesium for geographical information modeling.

Modeling tools are also model management tool for model evaluation, verification. From the view of model engineering, Zhang, Zhou and Horn [68] proposed a list of metrics to define the rightness of DT modeling from the lifecycle perspectives, including model construction, evaluation, and management. On the similar perspective, Tao *et al.*[69] performed multi-aspect analysis of the DT modeling via model construction, assembly, fusion, verification and modification.

Though the study also discussed other aspects of DT modeling, the modeling types, techniques and tools are considered primary properties to distinguish DT modeling.

### 2.2.6 Services

With the completion of the model of the twinning object, the final stage of DT implementation is to leverage the DT modeling and combining this with domain specific knowledge to provide service benefits to users. In this study, service benefits refer to the functions derived from purposes of digital twinning that fulfil the potential of the DT data and modeling.

Williams, Chatterjee and Rossi [70] identified a few fundamental design dimensions that distinguish one digital service from another, two of these dimensions are considered applicable to DT services - service resource and service delivery. From the angle of servitisation, Meierhofer *et al.*[71] considered DT as an enabler for servitisation of manufacturing hinged on its capability in the value creation. The service, created by data analytics or simulation is delivered to the system actors - production line, product or user. Both studies concern where the service is generated from and where the service is delivered to.

There are also comprehensive dimensions for DT services derived from software engineering. Qi, Tao and Nee [72] analysed the steps involved in the DT services running workflow: service request, resource collection, service encapsulation and service delivery. Aheleroff *et al.*[34] proposed Digital Twin as a Service (DTaaS) paradigm on the basis of Everything-as-a-service (XaaS) - a general category of applications enabled by cloud computing. DTaaS delivers four categories of DT services: transformation of data based on the DIKW hierarchy (data, information, knowledge, wisdom) model, integration of human workforce and cyberspace to achieve high level efficiency and accuracy, retrieving the semantic content of a number of assets and autonomous capacity to make an uncoerced decision via real-time connectivity.

It is concluded that the first main element of DT services is the resources that the service is generated from. The resources also implies the service maturity level, which can be rated by DIKW in terms of the extent of data value extraction. The DIKW model is applied due to its high consistency with the digital twinning purposes and processes. Based on the DIKW model, Qi *et al.*[73] summarised enabling technologies for DT services as platform services, resource services, knowledge services and application services. Hereby the maturity of DT services are classified as

- Data service where the presentation of data is the priority while processing data is set as minimum requirement. Examples are common data environment, data storage and retrieval, etc.
- Information services to provide semantics sourced from data analytics such as fault diagnostics, visualisation of current status of system, etc.
- knowledge service based on the construction of modeling for the purpose of problem investigation, examples are rule mining, decision tree, expert system, etc.
- Wisdom service based on the knowledge acquired of the system, with large-scale and in-depth analytics, such as simulation for system behavioural prediction under certain conditions.

The other key characteristic for DT service is service delivery which defines through which gateways the data and information leave the DT. Technologically, DT services can be divided into machine-to-human and machine-to-machine [15]. Machine-to-human services normally present outcome to users visually, such as augmented reality and dashboard. Machine-to-machine services deliver actionable intelligence to autonomous machines that can communicate with each other, transferring critical information required for asset operations[74].

# 3 Domain Specific DT Analysis

In this section, thematic analysis was conducted on literatures of each selected domain to identify archetypes of the domain specific DT use-cases. As a result, several collections of DT instances were listed. The description of DT instances follows the dimensions defined in the synthesised comparative framework, aiming to present a detailed outline of DTs in each domain, which were used as groundings for the cross-domain comparison.

## 3.1 Agriculture

As a process for cultivation of the soil to grow crops and the rearing of animals to provide food, agriculture DT could realise a sophisticated management system to maximise productivity while reducing labour requirement, energy usage and losses. A similar concept - precision agriculture, is found to be appeared earlier than agriculture DTs, both purposed on more timely understanding of farmland and livestock for optimised farming management.

The most discussed twinning object in the literatures is farmland which is a system of crops and environment. Services of agriculture DT emphasis on visible and automated management of irrigation scheduling, fertiliser application and detection of infectious disease [75]. Implementation of the system architecture tend to utilise established platforms, a typical example is an irrigation DT implemented by Alves *et al.* [76]. , where they applied low-cost sensors designed by Sensing Change, an IoT platform built in the SWAMP project and a data subscription service developed by FIWARE Foundation.

One distinctive characteristic of agriculture DTs is the direct involvement of living systems including animals [22] and plants. While unlike human DT in healthcare, agricultural processes tend to evolve relatively slow at the temporal dimension so high-frequency interaction between the physical object and DT is not vigorously required [35]. This feature can affect the technical development of agriculture DT, the data acquisition relies on IoT monitoring solutions featuring in low power and long range communication. The modeling of farming environment could be based on empirical equations [76] and services level mostly remains at sensing and basic analytics. The more advanced service, usually in a controlled environment where environment parameters are more manageable [77], is prediction and automation of nutrition application [78].

| Ref. | Object | Purpose | Architecture |
|---|---|---|---|
| [77] | Controlled environment of greenhouse | Optimise yields and quality of crops with energy consumption of greenhouse | Layered structure with functional hardware and software |
| [76] | Watershed's water balance and Irrigation of farmland | - Present different aspects and parameters that impact the farm's behaviour, yield production and resource consumption<br>- Enable farmers to make better decisions and to decrease the environmental impact in water, land and soil resources | Five layered reference model based on SWAMP project with various hardware and software tools |
| [78] | Field state and corp health condition of the farm | - Monitor soil parameters<br>- Automate the optimisation process of irrigation and fertilisation activities | Reference model showing logical connection of system elements |

| Ref. | | | |
|---|---|---|---|
| [22] | Environmental factors of a pig farm | - Discover out the best environmental conditions for growth<br>- Improve animal welfare and so minimise the unnecessary cost due to disease | Layered reference model based on commercial DT platforms (e.g. GE Predix, Eclipse Ditto, IBM Watson IoT) |

Table 2a: Agriculture Domain Digital Twin Instances under Comparative Framework - Part 1

| Ref. | Data | Model | Service |
|---|---|---|---|
| [77] | - Temperature and humidity sensors, operation status of exhaust fan and submersible pump.<br>- Data storage in MySQL at server. | - Data-driven modeling based on historical dataset in Energyplus.net<br>- Crop growing modeling based on Soil–Plant–Atmosphere dynamics in Dssat.net | - Python-based CDE for data processing<br>- Simulation of behaviour of heating and ventilation systems with Energyplus.net<br>- Simulation of growth and yield of crops for agricultural decision support with Dssat.net |
| [76] | Data collection via LPWAN sensors developed by Sensing Change to monitor soil, air and light. Including a raspberry PI based monitoring station and a smartphone application to view the real-time field data. | Watershed's water balance is modelled based on Penman–Monteith equation to calculate optimal soil moisture and control the irrigation based on environment. | - Iot data visualised on SWAMP environment<br>- Data queries and subscription services via FIWARE platform |
| [78] | - Image of plant leaves are captured by drone and uploaded to cloud server<br>- Soil parameters are measured by WSN and sent to cloud via LoRaWAN | - Images of plant leaves are processed by computer vision (e.g. MobileNet CNN) to detect disease and nutrient deficiency.<br>- Correlation of the result of the image processing is analysed with the data gathered from the WSN. | - Remote access to view the status of farmlands in near real-time<br>- Automated detection of crop diseases based on images Recommends optimal irrigation and fertilisation strategies |
| [22] | Sensors for measuring environmental factors (i.e., temperature, NH3, $CO_2$, humidity, dust, etc.), which can affect the growth of livestock | Optimal environmental condition is determined via big data model | - Simulation to suggest optimal temperature and $CO_2$ for livestock farms, and then operate fans and open windows as an execution<br>- Visualise data and analysis results in a user-friendly way using 2D/3D |

Table 2b: Agriculture Domain Digital Twin Instances under Comparative Framework - Part 2

## 3.2 Manufacturing

As a sector diligently pursuing production efficiency, manufacturing has been at the heart of industrial revolutions, starting from industrialisation to electrification, then automation and now digitalisation. Industry 4.0 has prepared numerous enablers for manufacturing DTs, such as Internet of Things (IoT), cloud computing, artificial intelligence, cyber-physical system, etc.

The digital twinning object in manufacturing domain could be any of the followings: product as unit level[79], production line as system level[80][81] and network of operational products as SoS level[21]. A product DT describes the geometrics, properties and function information of a product in the design, manufacturing and use phase to monitor its status over the whole life cycle, for the purpose of prognostic health management.

The integration of multiple unit-level DTs constitutes a system-level production DT, which could be a manufacturing line, a shop floor, or a factory. The purposes are optimising the allocation of manufacturing resources and improving production efficiency via semi-automation as human-robot-collaboration and total automation based on the closed loop of sensing-analysis-decision-execution [30]. The proposed system architectures have progressed to relatively mature level – technical architecture with architectural pattern.

Data collection benefit from the developed technologies - CPS and Industrial Internet of Things (IIoT), where data formats (e.g. STEP[53]) have been standardised. Blockchain also plays an important role by enabling the decentralised data storage of life-cycle product information[79]. The services for a product DT range from web-based visual simulation to present the construction of models at design stage, data dashboard at usage stage and predictive maintenance at operational stage.

| Reference | Object | Purpose | Architecture |
|---|---|---|---|
| [21] | Cyber-physical production system (CPPS) | - To achieve information symmetry between the CPPS and manufacturing employees<br>- Implement 'human-in-the-loop' perspective for better production performance. | Service-oriented architectural pattern Technological reference architecture |
| [79] | Life cycle operational data of all the manufactured turbine products. | To address the difficulty in management of product lifecycle data, as many participators constructing a complicated network with enormous data volume. | Service-oriented blockchain structured data management architecture |
| [80] | Operating status, including production elements and production processes, of the shop floor | - Continuously monitoring of the status of production process, shop-floor managers can make decisions timely.<br>- Accelerate response to production problem. | A layered functional shop-floor data management model is constructed to indicate data flow among system components |
| [81] | Robotic operation of a micro smart factory | - To solve inefficiency in the current Factory-as-a-service paradigm.<br>- Real-time monitoring of the present, tracking information from the past, and operational decision-making support for the future | Conceptual-level four layered interoperability-context system architecture. |

Table 3a: Manufacturing Domain Digital Twin Instances under Comparative Framework - Part 1

| Reference | Data | Model | Service |
|---|---|---|---|
| [21] | Data collection via IIoT in CoAP, data delivered to users in JASON | Ontology-based knowledge structure to map data generated by the CPPS. | - Augmented reality combined with a vocal interaction system to deliver manufacturing knowledge.<br>- Remote terminal units to serve as gateways to the knowledge model. |
| [79] | Sensor data indicating the dynamic product profile of the turbines are stored in a specific | Not included. | - The data management platform can be accessed through mobile device, to monitor the states of the turbine. |

| | block and chained in a peer-to-peer network | | - The entire blockchain can be presented on the platform, where a specific block can be explored through the search function.<br>- Performance optimisation and design improvements of the new turbine. |
|---|---|---|---|
| [80] | Location of logistics, equipment start and stop signal, motion data of equipment sent to the data centre. | Discrete events modeling (i.e. ESHLEP-N) builds the operation logic of shop-floor. Markov chain is used in the modeling of deduction rules. | - 3D virtual scene of the shop-floor is shown in Unity3D.<br>- Prediction of shop-floor operating status using Markov chain to assist managers to identify bottlenecks and optimise the production processes. |
| [81] | Information exchange in JSON format. RESTful API is used as the network architecture for the IIoT network layer | External (e.g. Mworks) robotics simulation engine. | Web-based communication environment for event handling and synchronisation. |

Table 3b: Manufacturing Domain Digital Twin Instances under Comparative Framework - Part 2

## 3.3 Construction

Digital twinning was found to present several applications in the design, construction, operation, and maintenance of an asset in the construction domain. Many researchers [57][82][83] proposed to combine DT and Building Information Modeling (BIM) which provides various digitised information of the physical assets, such as dimensions, material, structural connections, etc. While the twinning object of construction DT are the operational status. From the life cycle point of view, BIM mainly focuses on the design and construction phase of a structure, while construction DT are focused on the operation and maintenance phases, for the purpose of long-term monitoring and optimising the operational processes. Other purposes include reducing construction cost, improving quality and updating stakeholders with the latest information about the project.

The data of construction DT are mainly the status of the structure or facility which are acquired by either IoT sensors or reality capture technologies, such as laser scanning and photogrammetry. The modeling approach are normally a combination of the conventional structural model, BIM model and machine learning model, where BIM model is becoming the core of the system, so Industry Foundation Classes (IFC) is recommended by some studies[7][84] as the basis for the construction DT data exchange and storage. Construction comprises a wide range of high hazard activities, so addressing health and safety (H&S) is one of the main purposes of applying DT. AR and VR are critical DT service delivery technology to tackle the H&S issue, they can also improve the collaboration among multiple stakeholders[85].

The system architecture of DT use-cases are showing not only functional design details but also technical implementation tools, it is seen that efforts have been geared towards the adoption of DT in the construction sector, even though there is still gap between conceptual design and deployment.

| Reference | Object | Purpose | Architecture |
|---|---|---|---|
| [86] | Structural behaviour of a railway bridge | Monitoring the structural health of bridge<br>Creating a collaborative environment for stakeholders at various stages of the project | No provided |
| [85] | Construction environment and onsite-worker behaviour | Monitoring construction environment for safety purposes | Reference architecture displaying the workflow |
| [31] | Geometric and semantic information of the electrical and fire-safety equipment of the building | Acquiring information of the electrical and fire-safety equipment of the building | Reference model showing the components of the workflow |
| [87] | Decision analysis framework for the O&M OF tunnels | Guiding and optimising the O&M management | Layered and service-oriented architecture with technical implementation details |
| [56] | Operation and maintenance (O&M) of buildings | Predicting a building's O&M status and ensuring that the buildings work normally, as well as reducing the damage caused by functional errors. | Layered and component-based architecture with proposed implementation tools |
| [82] | Optimising maintenance processes and energy efficiency in ports | Developing energy-saving procedures and strategies and integrating production systems from Renewable Energy Systems (RESs) for sustainable mobility | Layered and service-oriented architecture with functional design details |

Table 4a: Construction Domain Digital Twin Instances under Comparative Framework - Part 1

| Reference | Data | Model | Service |
|---|---|---|---|
| [86] | Strain/stress data collected by fibre optic sensor networks | Integrating both physics-based finite element analysis model and data-driven machine learning model | Real-time sensor data and associated bridge behaviour are visualised in a BIM software. |
| [85] | Recording video and motion detection sensor | - Computer vision algorithm to detect unsafe factors and worker behaviours<br>- 4D BIM simulation for construction activity | - Issue warning to unsafe behaviours<br>- Record the occurrence of misconduct to generate knowledge base and training programs |
| [31] | Capture 2D information from images and 3D information from laser-scanned point clouds | - Semantic information on the devices is extracted by AI-based image segmentation<br>- Geometric of the devices are reconstructed using Structure-from-Motion (SfM) and Multi-View Stereo (MVS) software | The geometric and semantic information of electrical and fire-safety equipment are mapped to the 3D model of the building. |
| [87] | Physical data from real-time sensor monitoring and semantic data extracted from manual inspection, construction and maintenance activities in IFC format. | - Physical rule based structural model<br>- Knowledge retrieval model<br>- Visualisation model uses BIM | The extended COBie standard-based organisation, the semantic mapping-based ontological expression and the rule-based semantic reasoning of the tunnel |
| [56] | - Surrounding environment recorded by sensors such as cameras, humidity, smoke, etc.<br>- Building entity, stress sensors, strain sensors, n Equipment information such as water volume, electricity usage | Architectural structure model, building equipment model, energy consumption model, geometric model, physical model, machine learning(Neural Network) | Operating system development, status prediction, life prediction disease analysis and risk analysis |
| [82] | Wind and smart city data collect by IoT | Integration DT Model (BIM and GIS based) and LEDs | Providing energy saving strategies |

Table 4b: Construction Domain Digital Twin Instances under Comparative Framework - Part 2

## 3.4 Healthcare

The existing DT applications in healthcare domain primarily focused on precision medicine and healthcare services management. Twinning objects in precision medicine are mainly organs and human body for analysing and developing predictions in order to provide clinical advice for patients[88] [89]. While real-time supervision of healthcare services [90] aim to characterise health service delivery processes for effective demand management. The mostly proposed healthcare DT architecture reference model which means the DT applications generally stay at conceptual level.

DT for patient relies on wearable devices to collect data so wireless data communication (e.g. Bluetooth, Machine-to-Machine, etc) is typically required. Data privacy and security is comparably more sensitive in healthcare domain which means it may be hard to obtain the patient data for validation of modeling and prediction [91].

The DT modeling in healthcare domain share methodologies across disciplines, such as data-driven[89], discrete-time events [88] as well as agent-based [92]. While DT in this domain is believed to be more challenging due to the high level of uncertainty. For example, the mechanism of human behaviour in a socio-technical system like an ICU is largely unknown [90] which makes quantitative modeling of clinical decision-making difficult, thus approximate behavioural model is more feasible even it means fidelity may be compromised [91]. For the similar reason, full automation may not be favoured when consequences of system errors are non-tolerable, so human-in-the-loop is essential for services of healthcare DT [92].

| Reference | Object | Purpose | Architecture |
|---|---|---|---|
| [88] | Heart rhythms of patients | - Monitor health status continuously and early detect abnormal situations<br>- Enable healthcare professionals to prescribe the suitable treatments and test them in a safe environment | Reference model with functional data flow chart |
| [89] | Cardiology of patents | - Treatment and prevention of cardiovascular disease based on accurate predictions of underlying causes of disease | Reference model with conceptual system deign |
| [90] | Operation of ICU (Intensive Care Unit) | - Improve critical care delivery by effectively managing demand surge and alleviating physician burnout.<br>- Evaluation of ICU capacity and data generation during extreme scenarios.<br>- Assisting design of intervention strategies.<br>- Optimisation of ICU services aligned with priorities of all stakeholders. | Reference and functional architecture |
| [92] | Physiology of individual elderly patients and local medical resources | - Real-time supervision and accuracy of crisis warning for the elderly in healthcare services<br>- Predication and optimisation of medical resources for seasonal diseases. | Layered reference architecture compromising<br>- Healthcare resource layer<br>- Perception layer<br>- Virtualisation layer<br>- Service layer<br>- User interface layer<br>- Application and user layer |

Table 5a: Healthcare Domain Digital Twin Instances under Comparative Framework - Part 1

| Reference | Data | Model | Service |
|---|---|---|---|
| [88] | Heart rhythms captured by IoT wearable sensors and transferred to cloud database. | Data-driven classifiers and predictive models to detect anomalies and future conditions. | - Patents can access cloud database where the machine learning models results are stored.<br>- Healthcare professionals provide correction based on diagnosis.<br>- Healthcare professionals can compare similar cases for more advanced and accurate decisions. |
| [89] | Combined data resources from mobile health monitor, clinical reports, medical images | Combing induction using statistical models learnt from data, and deduction, through mechanistic modeling and simulation integrating multiscale knowledge and data | Guide clinical decision-making |
| [90] | Hospitalisation data, bed location data, medication data, IoT sensors monitoring the processes within the ICU. | A hybrid simulation model to simulate care delivery processes as discrete-time events, combined with behaviours of clinicians and patients in the same simulation environment to capture their interactions under a variety of ICU production conditions. | It is proposed that the services can be delivered by integrating the simulation with the hospital information system (e.g. EHR) |
| [92] | - Wearable monitoring equipment for real-time physiological data of elderly patients (e.g. blood pressure/oxygen, etc.)<br>- Digital healthcare records from healthcare institutions | - Dangerous events such as falling can be predicted through iteration of the virtual model using machine learning algorithms<br>- Iterative optimisation model to recommend dosage and frequency of medication<br>- Diseases incidence prediction model based on historical data for pre-arranged healthcare equipment and personnel | - Real-time supervision for medication reminder and health physiotherapy<br>- Crisis early warning (emergency, first-aid)<br>- Medical resource scheduling and optimisation (bed planning, clinicians allocation) |

Table 5b: Healthcare Domain Digital Twin Instances under Comparative Framework - Part 2

## 3.5 City

A city level DT is inclusive of several components including one or multiple systems such as transportation, environment and energy. When city-level DTs are integrated together, it is often referred as smart city. The smart city assembles individual city-level DTs and their interdependencies through a federated system, so that a coordinated approach for planning, predicting and managing city can be achieved. City digital twinning may be confused with construction DT due to the overlaps in their twinning objects, such as buildings and infrastructure. However, aspects of the twinning objects are diverted for the two domains, as considered by the authors. Construction DTs concern the design and building process of physical structures while city DTs focus on the socio-economic effects posed by urban infrastructure.

The purposes of city DT include monitoring the current state of the urban environment, rapid and effective response to emergencies, efficient assessment of design and planning, predicting situational development, etc. The value could be optimising use of resources, reducing service disruption, increasing resilience and raising the quality of life of citizens[93], but how to materialise such benefits through effective policy making remains a big challenge ahead[94].

Compared with other domains, city DT are more likely to relay on graphic visualisation data to acquire information as the sensor-based reality information is likely insufficient to provide dynamic spatiotemporal information about physical vulnerabilities[95], and the physical form of the city can represent the operational status of the twinning object in applications such as road traffic control. The visual data collected from city environment are fed to machine learning based computer vision models[96], so high quality of video transmission is required and cabled internet connection (e.g. broadband, fibre optics, etc) is normally a prerequisite for smart city DT, while there are also cases applying wireless telemetry if the volume of the data is relatively low, such as M2M communication for the purpose of harvesting energy usage data as well as billing customers by utility companies.

The referenced twinning objects of city DT are usually systems while it is widely believed that large scale, dynamic and highly complex city-wide DT is the ultimate goal [93], so it normally employs systems engineering principles. Agent-based simulation are seen in modeling of city DT for resources application where there is a large system consisting of autonomous and interacting individuals. Geographic information system (GIS) modeling is vigorously applied as the base layer for city DT where the topography, environment and spatial structure of the city are surveyed and mapped into a GIS-based database.

| Reference | Object | Purpose | Architecture |
|---|---|---|---|
| [34] | Water cycle of constructed wetland | To automate physical inspections including transportation to laboratories, providing reports and consolidating in a timely matter. | Service-oriented technical architecture |
| [97] | Traffic loads of bridges group in a regional transportation infrastructure network | Monitoring the traffic loads in physical bridge Evaluating working status of physical bridges | Reference model, consisting of two subsystems - hardware and software |
| [98] | Sustainable urban road planning | To provide a functional, economic, people-friendly, eco-friendly urban road planning scheme considering new road construction and existing old road widening to alleviate traffic congestion | Logical, functional and technical, layered and service-oriented |
| [95] | Distant objects in city that may lead to hazards in extreme weather events | Effective risk-informed decision-making for better disaster risk management | Logically structured flowchart showing the data and process |
| [82] | Energy generation system running status of wind farm | - To develop energy-saving procedures and strategies - To integrate production systems from Renewable Energy Systems (RESs) for sustainable mobility | Logical, functional and technical layered and service-oriented |
| [99] | Objects on the road that may affect the driving conditions of vehicles | To monitor the road conditions and enable self-driving function of vehicles | Flow chart indicating data interpretation process and technical-oriented platform setup |
| [100] | Flooding levels of rivers in the city and tidal levels near the coast | Two key issues dominate disaster management scenarios in Takamatsu City: 1) Increasingly flooding of local rivers caused by torrential rain and high tides along the coast and (2) the need to quickly assess shelter requirements when a disaster occur | Technical data flow chart centred on NEC's Data Utilisation platform service |

Table 6a: City Domain Digital Twin Instances under Comparative Framework - Part 1

| Reference | Data | Model | Service |
|---|---|---|---|
| [34] | Sensors to monitoring water level, water quality, and sediment | 3D model is developed based on GIS data, water top surface and sediment level are visualised based on sensor data. | Instant 3D in AR-enabled geographical viewer enriched with IoT data |
| [97] | Information fusion of weigh-in-motion (WIM) and multi-source heterogeneous machine vision | - Statistical models to analysis dynamic response spectrum of bridge<br>- Machine learning model (YOLO-v3 CNN) to identify traffic flow via machine vision | Issue safety warning when damages are detected |
| [98] | Geographic information, traffic information and environnemental information | Multi-criteria decision making and GIS | Focuses on interpreting various data from multiple sources in the physical world into a digital expression |
| [95] | Mapping and updating vulnerable objects relies on citizen reporting through 2D map-based enquiry or participatory sensing and crowdsourced visual data analytics | A model update based on unstructured crowdsourced visual data analytics to better understand the spatio-temporal information of physical vulnerabilities with respect to neighbouring critical infrastructure | Public can access interactive 3D visualisation in computer-aided virtual environment to view the vulnerable objects in their neighbourhood and likelihood of affecting critical infrastructure during extreme weather events. |
| [82] | Wind and smart city data collection by IoT | Integration DT Model (BIM and GIS based) and LEDs | - Provide energy saving strategies<br>- Optimising maintenance processes and energy efficiency in ports |
| [99] | An identification module (e.g. cameras) for recognition of vehicles and persons, including when and where an object appeared. A tracking module to identify the corresponding objects and keep track of all video frames where these objects appeared. | Machine learning models in JSON format (Single-Shot Detector and deep learning) model for car and person recognition, fused with GPS coordinated 3D road model. | - Citizen can view the 360° live streams of the road on web page or head-mounted devices for journey planning.<br>- Authorities can be alerted of dangerous object and generate automatic statistical reports of human/car circulating the area to optimise traffic planning |
| [100] | Real-time checking of water and tide level sensors at the observation points located throughout the city. Rainfall data provided on weather forecast website. | Not introduced. | - Visualisation of disaster management data<br>- Services developed based on FIWARE that can provide freely access and use of public data service to citizens and businesses and government. |

Table 6b: City Domain Digital Twin Instances under Comparative Framework - Part 2

# 4 Cross-Domain Digital Twin Comparison and Solutions

As illustrated in Figure 5, this section of study starts from the observed 'surface-level' similarities and differences from domain-specific DTs, then commits to investigate 'deeper-level' correlations that assimilate or differentiate the DTs of different domains, to provide explanations on why DTs are the way they are. Finally, based on the commonalities shared across the domains, unified approaches to conceptualise and implement DTs are proposed, where the differences of a variety of DT instances are being encompassed in the cross-domains solutions.

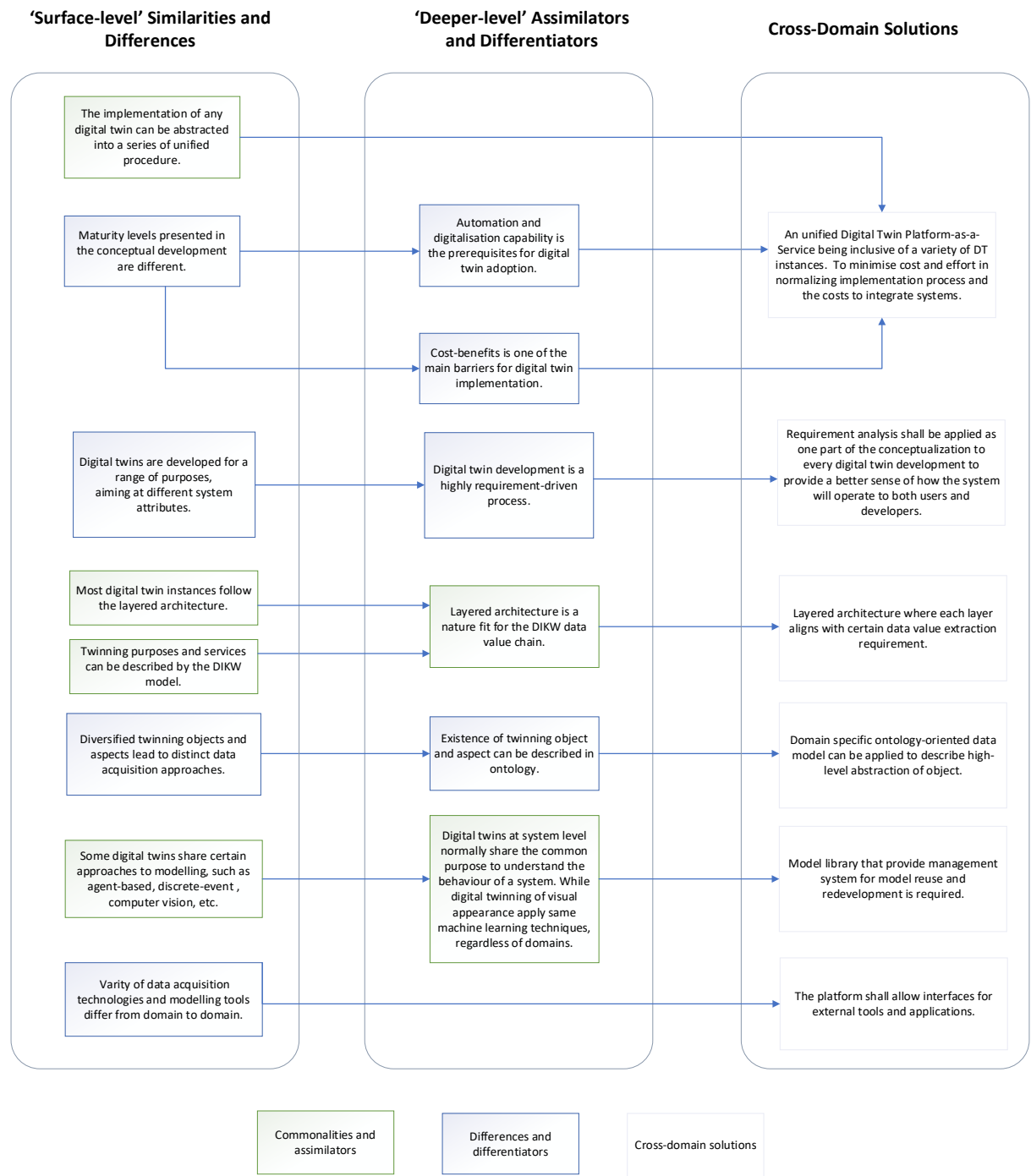

Figure 5: Derivation of Cross-Domains DT Solutions

## 4.1 Requirement-Driven Digital Twin Conceptualisation

It is noted from Section 3 that the purposes of DTs within different domains are common, including real-time monitoring, integration tool, fault analysis, prediction, optimisation, etc. The classification of the purposes is mainly based on the level of capability. Though there are potentially more benefits and values that the DT could provide by 'climbing up' the DIKW hierarchy, there are also increases in system complexity and so the cost and effort to build such systems[101]. Therefore, instead of pursuing the higher levels of maturity and sophistication, the DT users and developers shall develop the DT to the levels that are fitted to the specific needs and purposes.

With the guidance of the comparative framework described in Section 2.2, a set of requirements for a DT framework can be derived from the object conceptual model, twinning purpose and system architecture. The object conceptual model may include any design related to the nature of the object, such as the federation and aspect of the object which suggest the modeling technique to be applied[4]. The purpose specification identifying the demand and value shall lay foundations and define boundaries for developing future DTs[19]. The architecture specification is closely linked to the system quality attribute such as re-usability, extensibility, interchangeability and maintainability across the entire DT lifecycle[32]. The requirement developed from the conceptualisation process set the baseline framework and all the aspects of these requirements shall be addressed in the DT development[102].

## 4.2 Reusable and Interoperable Data Model and Modeling Library

The case studies in Section 3 reveal that neither raw DT data nor modeling data format follow universal standards. Raw data format is normally determined from data source, such as unstructured (e.g. sensor data), semi-structured (e.g. JSON) and structured (e.g. spreadsheets). While modeling data are rather domain dependent - IFC for construction sector, GML for city domain and STEP/AutomationML for manufacturing application. Though standardising all the DT data might not be practical, it is still worth investigating a solution for the consistency issue in the DT integration.

It is noticed that raw data type and format are highly dependent on types of sensors, actuators and communication requirement. It is analysed that the potential for DT standardisation exist in the cyber layer. There are two commonly applicable process/tools identified: 1) an intermediate data model that converts raw data from various data sources to a standard format that the DT engines could understand; 2) generalised commonly applicable DT models (e.g. agent-based, discrete event, image processing, etc) which are scalable and extensible by users according to services required.

From the case studies in Section 3, it can be seen that the interested twinning aspect are normally the operation status of the object, judged by the appearance captured by camera/video, or performance indicators recorded by sensors. From the principle of universality, data models which contain commonly interested properties such as object behaviours could facilitate the model reuse. The data model could be object-oriented as it describes ontology of the object. DT developer could select the particular aspects required for digital twinning to store semantics of data and support modeling process [103].

Similarly, a DT model library can be applied to collect a series of reusable models so that users can match appropriate models according to their twinning purposes and further develop the model details to meet the requirement of targeted services[104].

Some efforts have been carried out in DT data model and reusable model library. For example, the Smart Data Model and the Foundation Data Model discussed in Section 2.2.4 exhibit interoperability and reusability. Industry Data Models by IBM [27] are consisting of a set of organisational and technical data models that are predesigned to meet the needs of a particular industry. A conceptual model reuse solution based on semantic mediation is discussed by Shani and Broodney [105]. Boyle and

Mackay[106] designed a discrete event simulation model to be reusable for a range of health service allocation. As an implemented solution, a Python-written agent-based modeling framework named Mesa has been demonstrated[107].

## 4.3 Digital Twin Services- Climbing the DIKW Value Chain

Predominantly, all DTs system could be considered as layered-architecture by its definition[73] - physical space, digital space and connections between the two are the three basic layers for creating a DT system. The physical space may also include edge layer for local computing demand, whilst the cyber/digital layer could contain cloud layer and application layer for data storage, processing, and functionalities. Particularly, the cyber space architectural patterns can be further differentiated according to the intended services, such as event-driven, service-oriented, big data-led, etc.

The categorisation of architecture summarised in 2.2.3, namely reference model(logical), reference architecture (functional) and software architecture (technical) can be considered as a key indicator for maturity.

On the time scale, though publications before 2020 propose simple form of architecture (i.e. reference model), there are more publications after 2020 describing reference architecture which maps system elements to functionalities, this can be considered as a step-up in the maturity of implementation.

On the domain spectrum, there are more mature architectural solutions in manufacturing, supported by larger percentage of reference architecture compared with reference models. City DTs, as system of systems, require compatibility to achieve interoperability of sub-systems, and also scalability for addition of more and more sub-systems progressively. Cities DT can be also service-oriented. As for healthcare DTs, the medical system DTs are mainly event-driven in order to be responsive.

The Gemini Principles set the purpose of DTs is to provide determinable insight into the built environment. It is noted that the DT architecture and services are structured following the DIKW model, a hierarchical framework that can enable extraction of insights and value from data, regardless of the domains. As a result, the services/applications generated by the DT system could be classified based on the DIKW model, as described in the Section 2.2.6.

| DIKW Level | Reflection on Six-Dimensional Framework | | |
|---|---|---|---|
| | **Purposes** | **Architecture** | **Services** |
| **Data** | Passive monitoring | Data acquisition in physical space | - Common data environment<br>- Data storage and retrieval |
| **Information** | Reactive analytics | Semantic information extracted from data via data engine | - Visualisation and/or notification of current system status<br>- Fault diagnostics |
| **Knowledge** | Prediction of future status | Modeling engine to assist investigation of reasonings of certain system behaviours | - System behaviour modeling to provide reasoning via ontologies, etc.<br>- Expert system that uses databases of expert knowledge to offer advice |
| **Wisdom** | Proactive management | Service engine to generate user-required service based on data, | - Prediction of system behaviour on the basis of data and models<br>- Decision-making based on multi-objective optimisation |

| | | information and knowledge | |

Table 7: DIKW-based Digital Twin Architecture and Services

## 4.4 Differentiators for Readiness Levels and Perspectives

Although DTs of various sectors share many similarities, there are several observed inconsistences. From the implementation perspective, it can be easily understood that due to differences in twining objects and required twinning aspect, data acquisition technologies and modeling tools are distinctive.

From the conceptualisation perspective, the sheer difference noticed are the purposes of digital twinning, which is also the fundament defined in Gemini Principles[19]. Primary industry (i.e. agriculture) extracts products from nature, so monitoring the production environment is priority. Secondary sectors (e.g. manufacturing) are similar production process but in a more controlled environment (e.g. factory, shop floor, etc.) which lead to the demand for smarter management and autonomous operation. Construction domain concerns full life cycle of buildings from design, construction to operation, therefore both real-time services and optimisation services are important. Healthcare focus on more precise and personalised services, while city DTs focus on urban governance and cross-departmental collaboration. The variations come from nature of value creation of sectors.

Nevertheless, from the view of twinning object, each domain requires DTs from unit level to system level, unit DT concerns the status of the unit itself while system DT focus on maintaining and improving overall system performance. It is argued that the observed differences at conceptual level are related to the maturity development to some extent.

In the following sections, three differentiators that lead to the current readiness level and future development of DTs are identified and discussed.

### 4.4.1 Digitalisation Capability and Controllability

Within the context of autonomous DT, the process starts from converting physical objects into digital models that could be recognised by computers. Then the computer shall utilise the digital models to predict and optimise the future status and drive actuators to intervene the physical object based on the results of the optimisation. There are two procedures – digitising and intervening the physical object, that are closely related to the twinning object. So it is analysed that the ability to implement the two procedures – digitalisation and controllability are the one of the main differentiators of readiness level of DTs for each domain.

In domains where twinning objects are relatively static and can be modelled by a small number of parameters, implementation of DT are comparably easier than domains where twinning objects require modeling by high-dimensional data, because the large number of features that may affect the future status of the object add-up the difficulty in prediction. For example, modeling the operation of a machine only require few parameters, while human body with the constant molecular and physiological changes means extracting precise medical data as well as modeling of body condition is very difficult [108]. On the system-level DTs, if the casual links in the environment are directly driven by human activities, the digitalisation capability is promising, such as built environment, greenhouse, etc. On the other hand, when the process of the system is exposed to uncontrollable factors, it is much more challenging to implementing a DT, such as natural environment [109].

On the other hand, controllability also play an important role in DT readiness level. Enders and Hoßbach[9] suggested that production machines and products in manufacturing domain already have sufficient prerequisites for the use of DTs – bi-directional and automated connection of production machines to industrial IoT, so real-time or near real-time control of machinery is already enabled. On the contrast, construction DT may require a much more complex management tool acting as connection between the physical object and DT, this management tool largely rely on human to make decision and take action to change the physical object [110].

Digitalisation capability and controllability are also linked to the misrepresentation of DTs, as the small errors can be amplified under cascaded interactions and interactive algorithms [111], especially in domains where the twining objects are system or system of systems.

### 4.4.2 Cost-Benefit Weighing

Cost-benefit is considered as another differentiator for the current DT readiness level. The deployment of sensors, communication network and software platform require upfront investment, while availability of investments and digital infrastructure differ from organisations and sectors.

Integration level, Interval (details and accuracy) and Intricacy (resource required) should be assessed to weigh the trade-off between the business value and cost of creating DT [34]. Madubuike et al. [7] suggested that the massive scale of building DT and the volatile nature of the sector make adopting DT in the construction industry a difficult task.

The difference in the sector ecosystem and organisational structure can also affect the investment decision on implementation and deployment of DTs. Some domains like agriculture, manufacturing, healthcare may not involve much inter-party collaboration. While construction and city domains require participation from multiple parties throughout developing, operating and maintaining phases of DTs. The upfront investment in creating a DT during the design and implementation phases may not directly benefit those responsible for collecting operational data. Thus, the collective agreement on implementing DT is more difficult to reach.

### 4.4.3 Socio-Ethical Risks

Besides the technical and economic differentiators, unique socio-ethical challenges may become barriers and restriction to the adoption of DTs in some domains.

In labour-intensive sectors, psychological barrier of key producers could deter the implementation of DT. For agriculture domain, the use of DT in small-scale farm could not be welcomed by low-skilled workers [112]. To act alongside complex technological and manufacturing systems, new competencies are required of industrial workers which means additional training and learning[113]. These are the barriers to overcome while the impact is considered short-term and not significant.

Robust governance mechanisms are considered essential for DT application on precision healthcare, due to the ethical concerns of data mining on lives of human [114] and inequality caused by the inaccessibility and unaffordability [115]. It is believed that city-scale DTs, which uphold private data of individuals and potentially affects decisions on governance, shall withhold security standards[4] and exhibit transparency and accountability [111] before deployment. Generally, DT domains where data of human are collected and lives of human are impacted, are associated with socio-ethical risks and these risks could attract regulatory restrictions which may be one of the causes to slow down the DT advancement.

## 4.5 A Unified Solution - Digital Twin Platform-as-a-Service (DT-PaaS)

It is noticed that most DTs described in academic publications are on conceptual or prototyping level. To successfully implement a DT, the developer team must not only understand user requirement and domain-specific knowledge, but also being skilful with a variety of new ICT technologies including Internet of Things, web development and machine learning.

The adoption of DTs is facing some domain-specific challenges and the conceptualisation and implementation of DTs across different sectors are sharing the linking principles. Hence, there is potential to standardise the process and system components and integrate them into a unified framework to assist researchers and practitioners developing DTs suitable to their needs.  The solutions developed in investment-rich domains could be transferred to those with low investment incentives area.

Another incentive for a cross-domain DT platform is DTs of different sectors need to be interconnected for data sharing of global issues such as pandemic and global warming.  An example is the National Digital Twin programme initiated by CDBB where a vision for an ecosystem of connected DTs in formed by including building DT, transport DT and healthcare DT to achieve savings for stakeholders and ensure societal benefits for all[60].

Some early adopters of DTs have been able to leverage their expertise to provide software and platforms, such as GE Predix, Bentley iTwin and Microsoft Azure DT. While the current commercialised solutions pose following limitations: 1) Modeling types are limited to geometrics and GIS, no mechanism-based modeling or interface for any mechanism-based modeling are provided, no data-driven modeling methods are integrated; 2) The available services stay at data visualisation and some basic semantic information service, such as GE Predix which could provide alarm when a sensor reading is above a threshold value. Azure DT could allow creation of standardised object-oriented ontology in JSON format and link it to telemetry data[116]. While Microsoft Azure DT is showing sign of great potential with openness enabled by CLI(Command Line Interface), which could interface with IoT data input, modeling tools and service delivery hardware (e.g. mixed reality headset).

Nevertheless, the current DT platforms are basically integrating IoT services with GIS/CAD modeling, while openness and interoperability can be further improved. With more and more open-source modeling tools, it is predictable that these DT platforms could involve to more compatible DT environment.

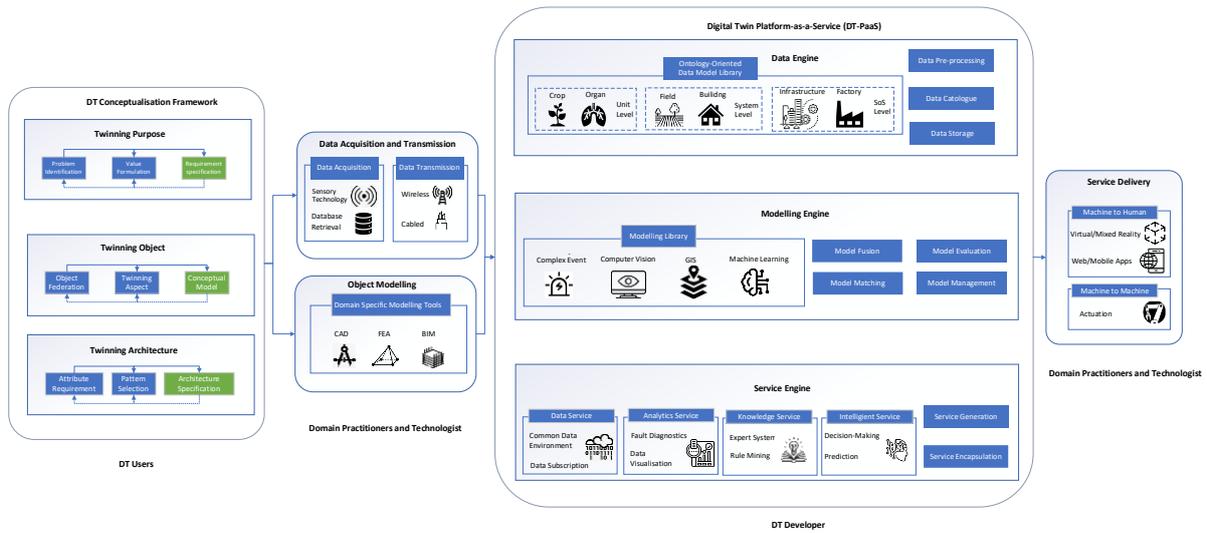

Figure 6: Proposed Cross-Domain DT-PaaS

Based on the DT features concluded in the six-dimensional comparative framework and the differences and similarities observed in the comparative study, a cross-domain DT development solution built upon data models, modeling library and various service applications is illustrated in Figure 6.

To begin with, perspective DT users shall go through the conceptualisation process to clarify the requirement specification of the twinning purposes, conceptual model of the twinning object and system architecture specification.

As the next step, the summarised requirements shall be discussed with domain practitioners and developed into practical data and modeling solutions. Then the application-specific data and modeling solutions shall be fed into the corresponding engines on the DT-PaaS platform, where data models and modeling libraries have methodised for reusability and interoperability for different DT instances. With the standardised object data models, an eco-system where multiple DTs interoperate with each other could be facilitated by using the ontology-oriented semantic data model from each DTs as input to cross-domain analytic process.

The DT-PaaS platform can be linked with other software via APIs, for example modeling tools and hardware for sensing and service delivery.

The authors believe that the technological barriers between domains can be broken by digitalisation and DT. A DT developer will be able to work in any domain with regards to the manipulation on domain-specific data, modeling and services. However, contribution from domain practitioners are still required when it comes to the interfaces with the physical object, such as data acquisition, mechanisms based simulation, interaction and control with actuation, etc.

# 5 Conclusion

A wide scope literature review was carried out in this study, including not only academic publications but also industry reports from relevant companies and organisations (i.e. FIWARE, Siemens, Digital Twin Consortium, Centre for Digital Built Britain). All aspects of DT keep evolving. A method or practice is urgently needed to identify and improve common rules that exist within a wide variety of DT systems. The three main contributions of this paper to the DT research community are as followings:

1. On the grounds of recognised research and engineering principles, a six-dimensional DT framework was proposed which follows the universalised creation process of DT from any domain. This framework can not only describe any DT but also used as a metric system for comparison.
2. The applicability of the proposed framework was demonstrated with various use-cases of DT. The similarities and differences of these DTs were summarised and analysed to provide explanatory theory to explain why DTs of different domains are the way they are.
3. A paradigm for Digital Twin Platform-as-a-Service (DT-PaaS) is deduced from cross-domain solutions that can integrate the universalised common process and tools, meanwhile allow the variations to be addressed via open interfaces.

The comparative framework in Figure.2 provides a concise guide on how a DT can be developed from raw ideas by identifying and contrasting the major methods, approaches, tools and technologies that may be involved in the conceptualisation and implementation of a DT, regardless of the domain. As a deeper study shown in Figure 4, this paper then investigates differentiators and assimilators that shape the DTs thus establishes an explanatory theory that uncovers the drivers, requirement, workflow of developing a DT concept and implementing the concept into a real-world application.

This study contributes to the paradigm of DT in future research and practice. Through framing future DT use-cases and destructing DT applications within a common cross-domain framework, the knowledge of DT can be stored, analysed and shared across domains, thus potential of DT can be fully realised and get the best out of DT capabilities in response to the digitalisation transformation.

DT research and practice are taking place across various sectors. It is expected that more diverse empirical evidence could be assembled to establish a shared body of knowledge and universal pipeline for leveraging and regulating the power of digital twinning.

**Acknowledgements**

This research did not receive any specific grant from funding agencies in the public, commercial, or not-for-profit sectors.